\documentclass[%
 reprint,
 amsmath,amssymb,
 aps,
 superscriptaddress
]{revtex4-2}
\usepackage{graphicx}
\usepackage{dcolumn}
\usepackage{bm}
\usepackage{comment}
\usepackage{physics}
\usepackage{lipsum}
\usepackage{siunitx}
\usepackage{makecell}
\usepackage{xcolor}

\begin{document}

\title{Harnessing micro-Fabry-Perot reference cavities in photonic integrated circuits}

\author{Haotian Cheng}
\affiliation{Department of Applied Physics, Yale University, New Haven, CT 06520, USA}

\author{Chao Xiang}
\affiliation{Department of Electrical and Computer Engineering, University of California, Santa Barbara, Santa Barbara, CA 93106, USA}
\affiliation{Department of Electrical and Electronic Engineering, University of Hong Kong, Hong Kong, China}

\author{Naijun Jin}
\affiliation{Department of Applied Physics, Yale University, New Haven, CT 06520, USA}

\author{Igor Kudelin}
\affiliation{Electrical Computer \& Energy Engineering, University of Colorado, Boulder, CO 80309, USA}
\affiliation{Department of Physics, University of Colorado Boulder, 440 UCB Boulder, CO 80309, USA}
\affiliation{National Institute of Standards and Technology, 325 Broadway, Boulder, CO 80305, USA}

\author{Joel Guo}
\affiliation{Department of Electrical and Computer Engineering, University of California, Santa Barbara, Santa Barbara, CA 93106, USA}

\author{Matthew Heyrich}
\affiliation{Electrical Computer \& Energy Engineering, University of Colorado, Boulder, CO 80309, USA}
\affiliation{Department of Physics, University of Colorado Boulder, 440 UCB Boulder, CO 80309, USA}

\author{Yifan Liu}
\affiliation{Department of Physics, University of Colorado Boulder, 440 UCB Boulder, CO 80309, USA}
\affiliation{National Institute of Standards and Technology, 325 Broadway, Boulder, CO 80305, USA}

\author{Jonathan Peters}
\affiliation{Department of Electrical and Computer Engineering, University of California, Santa Barbara, Santa Barbara, CA 93106, USA}

\author{Qing-Xin Ji}
\affiliation{T. J. Watson Laboratory of Applied Physics, California Institute of Technology, Pasadena, CA 91125, USA}

\author{Yishu Zhou}
\affiliation{Department of Applied Physics, Yale University, New Haven, CT 06520, USA}

\author{Kerry J. Vahala}
\affiliation{T. J. Watson Laboratory of Applied Physics, California Institute of Technology, Pasadena, CA 91125, USA}

\author{Franklyn Quinlan}
\affiliation{National Institute of Standards and Technology, 325 Broadway, Boulder, CO 80305, USA}
\affiliation{Electrical Computer \& Energy Engineering, University of Colorado, Boulder, CO 80309, USA}

\author{Scott A. Diddams}
\affiliation{Electrical Computer \& Energy Engineering, University of Colorado, Boulder, CO 80309, USA}
\affiliation{Department of Physics, University of Colorado Boulder, 440 UCB Boulder, CO 80309, USA}
\affiliation{National Institute of Standards and Technology, 325 Broadway, Boulder, CO 80305, USA}

\author{John E. Bowers}
\affiliation{Department of Electrical and Computer Engineering, University of California, Santa Barbara, Santa Barbara, CA 93106, USA}

\author{Peter T. Rakich}
\affiliation{Department of Applied Physics, Yale University, New Haven, CT 06520, USA}

\begin{abstract}

Compact photonic systems that offer high frequency stability and low noise are of increasing importance to applications in precision metrology, quantum computing, communication, and advanced sensing technologies. However, on-chip resonators comprised of dielectrics cannot match the frequency stability and noise characteristics of Fabry-Perot cavities, whose electromagnetic modes live almost entirely in vacuum. In this study, we present a novel strategy to interface micro-fabricated Fabry-Perot cavities with photonic integrated circuits to realize compact, high-performance integrated systems. Using this new integration approach, we demonstrate self-injection locking of an on-chip laser to a milimeter-scale vacuum-gap Fabry-Perot using a circuit interface that transforms the reflected cavity response to enable efficient feedback to the laser. This system achieves a phase noise of -97 dBc/Hz at 10 kHz offset frequency, a fractional frequency stability of $5 \times 10^{-13}$ at 10 ms, a 150 Hz $1/\pi$ integral linewidth, and a 35 mHz fundamental linewidth. We also present a complementary integration strategy that utilizes a vertical emission grating coupler and a back-reflection cancellation circuit to realize a fully co-integrated module that effectively redirects the reflected signals and isolates back-reflections with a 10 dB suppression ratio, readily adaptable for on-chip PDH locking. Together, these demonstrations significantly enhance the precision and functionality of RF photonic systems, paving the way for continued advancements in photonic applications.
\end{abstract}

\maketitle

\section{\label{sec:intro} Introduction}

Fabry-Perot (FP) resonators have many unique properties that make them indispensable for applications ranging from ultra-stable lasers~\cite{kessler2012sub,guo2022chip}, optical clocks~\cite{ludlow2015optical}, microwave signal generators~\cite{liu2024low,kudelin2024photonic}, to quantum networks~\cite{kuhn2002deterministic}.
Since they host electromagnetic modes that live almost entirely in vacuum, vacuum gap FP cavities exhibit greatly reduced frequency instabilities relative to dielectric resonators~\cite{liang2010whispering,kessler2012sub,jin2022microfabrication}.
To enable next-generation quantum communications, computation, and timekeeping technologies, it will be necessary to bring the performance advantages of FP resonators to compact, integrated platforms~\cite{kudelin2024photonic} (Fig.~1(a)).
As the basis for integrated technologies, new techniques for wafer-scale fabrication of vacuum-gap micro-Fabry-Perot (\textmu FP) cavities have produced compact reference cavities with ultrahigh Q-factors ($>\!10^9$) and excellent frequency stability~\cite{jin2022micro,liu2024ultrastable}.
However, before we can harness these performance advantages in next-generation integrated photonic systems, new strategies are needed to interface \textmu FP reference cavities with photonic integrated circuits (PICs). 

The most challenging obstacle that prevents the introduction of \textmu FP into PICs is the born nature of the wide-band and near-unity reflection of FP cavities when off-resonance. Due to the lack of on-chip circulators, such a strong reflection will destabilize and even damage on-chip lasers/ amplifiers. To enable the use of \textmu FP cavities as on-chip frequency references, we require efficient access to the resonator modes and methods to transform the cavity reflective response for various applications. For instance, electrical feedback can be used to actively stabilize a laser to the modes of an FP cavity by detecting the reflection response using Pound-Drever-Hall (PDH)-based feedback control.  
However, such schemes require the protection of the on-chip laser from back-reflections and efficient signal redirection for detection. 
To address this challenge, a novel reflection cancellation circuit has been developed that adapts conventional schemes for poor man’s isolator into a circuit architecture~\cite{cheng2023novel}. 
Alternatively, the frequency stability of a \textmu FP reference cavity can be transferred to a laser oscillator using optical feedback through self-injection locking~\cite{kondratiev2017self}. 
This method eliminates complex feedback control while producing noise suppression at higher bandwidths using a simpler system architecture. 
However, efficient and stable self-injection locking would require new strategies to shape and control the response of the \textmu FP cavity to enable resonant feedback with integrated laser~\cite{tkach1986regimes}.
Hence, innovative circuit interfaces could facilitate a new generation of ultra-high-performance lasers and oscillators by bringing the performance advantages of vacuum-gap reference cavities to integrated photonics.

In this paper, we introduce novel reflection transformation circuits~(RTC) for co-integration of vacuum-gap \textmu FP reference cavities with photonic integrated circuits to enable compact and ultra-high-performance lasers and oscillators. 
To harness \textmu FP cavities for self-injection locking, we demonstrate a new RTC that transforms the reflection response of the \textmu FP to produce the requisite resonant feedback; using this system, we demonstrate self-injection locking of a waveguide-integrated laser to an in-vacuum bonded \textmu FP reference cavity that was fabricated using a wafer-scale process. 
Self-injection locking of the DFB laser to the \textmu FP reference cavity, we achieve a single side-band phase noise of -97 dBc/Hz at 10 kHz offset frequency, $5\times10^{-13}$ fractional frequency stability at 10~ms, with a 150~Hz $1/\pi$ integral linewidth and a 35~mHz fundamental linewidth, corresponding to record performance of a self-injection locked laser within a waveguide-integrated system. 
To harness the frequency stability of \textmu FP reference cavities using PDH-locking schemes, we demonstrate an alternative approach for fully integrating a \textmu FP cavity using a complementary RTC~\cite{cheng2023novel}.
This integrated \textmu FP is used to demonstrate an isolator-free circuit interface for PDH locking, consisting of a circuit-based two-port interferometer that cancels unwanted back-reflections while mapping the cavity's reflection response to a separate optical port.
Looking ahead, these approaches to harnessing reference cavities within integrated circuits open new avenues for ultra-high performance portable photonic microwave systems~\cite{kudelin2024photonic}, integrated optical gyroscopes~\cite{gundavarapu2019sub}, and enhanced quantum systems~\cite{lai2022ultra} with the continued advancement of such \textmu FP reference cavity technologies.

\begin{figure*}[!t]
    \includegraphics{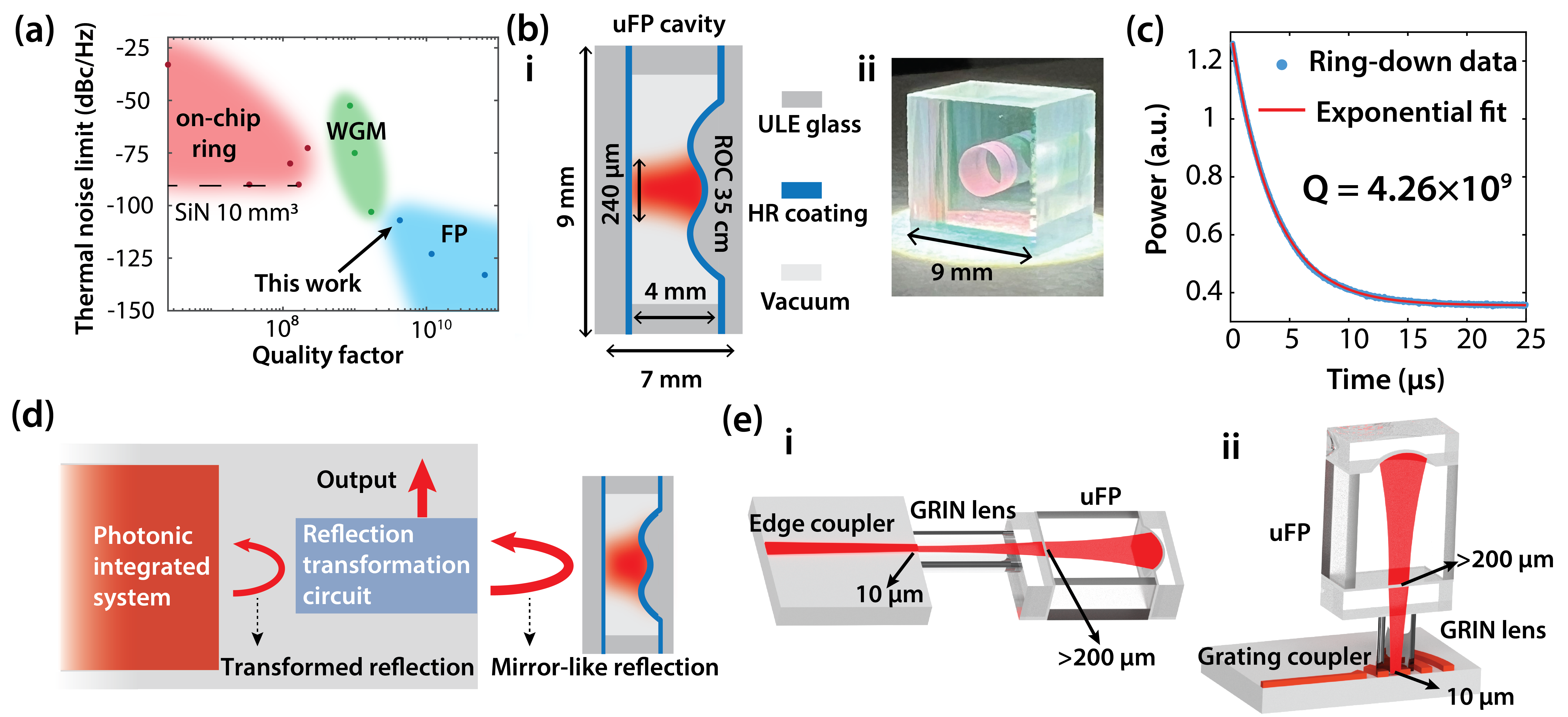}
    \caption{\label{fig1} \textbf{Integration of \textmu FP cavitiy with PICs. }
    \textbf{a}, Survey of resonator thermal noise limit at 10~kHz and quality factor for on-chip ring cavity, WGM cavity and vacuum FP cavity. Ring resonator data is from \cite{jin2021hertz,li2021reaching,sun2024integrated,idjadi2024modulation,he2024chip}, WGM resonator data is from \cite{liang2010whispering,wu2020greater,jin2024microresonator}, and vacuum FP resonator data is from this work and \cite{kessler2012sub,guo2022chip}. Note that our cavity has $> 10^4$ times smaller form factor than \cite{kessler2012sub} and doesn't require vacuum enclosure.
    \textbf{b}, \textbf{i}, Schematic illustration of the in-vacuum bonded \textmu FP cavity. ROC: radius of curvature. \textbf{ii} Photo of \textmu FP cavity.
    \textbf{c}, Optical ring-down measurement of \textmu FP cavity yielding quality factor of 4.26 billion.
    \textbf{d} Schematic of the integration architecture. A reflection transformation circuit is required to transform mirror-like reflection to the response desired by application.
    \textbf{e} Schematic of mode-matching approach from PIC to \textmu FP with \textbf{i} end-fire and \textbf{ii} grating coupler.
    }
\end{figure*}

\section{\label{sec:Co-packageFP} co-integration of \textmu FP cavities}

Integration of high-performance vacuum-gap \textmu FP reference cavities with waveguide-based PIC platforms presents several challenges. When directly probed, an FP reference cavity will reflect laser light back to its source, which is usually unfavorable for stable laser operation. In this work, we have developed a general on-chip interferometric platform, named reflection transformation circuit (RTC),  that can address this problem (Fig.~1(d). Based on the design, such a platform can either reshape the spectral response of the FP reference cavity reflection that enables on-chip laser self-injection locking, or completely redirect the reflection signal without the need of a circulator.
The second challenge is the large difference in size between the modes supported by waveguides and FP reference cavities, where the latter are usually an order of magnitude larger. To overcome this, we demonstrate two complementary strategies with edge and surface couplers to efficiently interface and access millimeter-scale reference cavity modes from micron-scale waveguide modes in two steps.
An inverse (Fig.~1(e,i)) taper or grating coupler (Fig.~1(e,ii)) is first used to expand the waveguide mode to several microns.
A micro-optic gradient-index (GRIN) lens is then used to expand and collimate the beam to a mode field diameter of 240 \textmu m, as necessary for efficient mode-matching with the \textmu FP cavity modes.

The vacuum-gap \textmu FP reference cavity used for these studies is shown in Fig. 1(b, i-ii), and we use a wafer-scale fabrication process to create such a cavity. A chemically assisted reflow process is used to produce an array of concave micro-mirrors with 35~cm radius of curvature on a super-polished 2-inch glass wafer~\cite{jin2022microfabrication}. Through optical-contact bonding of the concave and flat mirror wafers onto a matched spacer, we construct a \textmu FP cavity array that can be diced into individual cavities (Fig.~1(c,ii)). To eliminate unwanted sources of noise and frequency drift, the mirror and spacer are fabricated from ultra-low expansion glass, and the cavity assembly is bonded in a vacuum environment. The resulted plano-concave \textmu FP reference cavity has compact dimensions of $9\times 9\times 7$ mm, a small volume less than 0.6~mL and a Gaussian mode field diameter of 240 \textmu m.  Using optical ring-down spectroscopy, we measured the optical quality factors of these cavity modes to be $4.26\times 10^9$ (Fig. 1(c)), corresponding to a cavity linewidth of 45~kHz and a finesse of $8.25\times 10^5$. Through independent measurements, such vacuum-gap \textmu FP reference cavities have been shown to produce excellent phase noise and frequency stability~\cite{liu2024ultrastable}. The fabricated \textmu FP cavity with hermetically sealed vacuum gap offers not only unparalleled noise performance free from the limitations of thermo-refractive noise in dielectric cavities (Fig.~1(a)) but also eliminates the need for a vacuum enclosure. Hence, the integration of such high-performance \textmu FP cavities could bring these benefits to PIC platforms to enable next-generation compact lasers, oscillators, and sensors.

We use a chip-based RTC and \textmu FP reference cavity to explore two distinct applications for optical feedback and electrical feedback.
We first demonstrate direct self-injection locking of an on-chip laser to a \textmu FP reference cavity in Section~\ref{subsec:End-fire}. 
This method harnesses the passive optical feedback to enhance laser stability directly with the \textmu FP cavity.
For the grating coupler based demonstration, presented in Section~\ref{subsec:Grating coupler}, we employ the technique~\cite{cheng2023novel} for \textmu FP signal redirection and back-reflection cancellation within the co-integrated platform. This configuration facilitates the effective use of the \textmu FP cavity for electrical feedback in a more general RF photonic platform.
\begin{figure*}[hbt!]
    \includegraphics{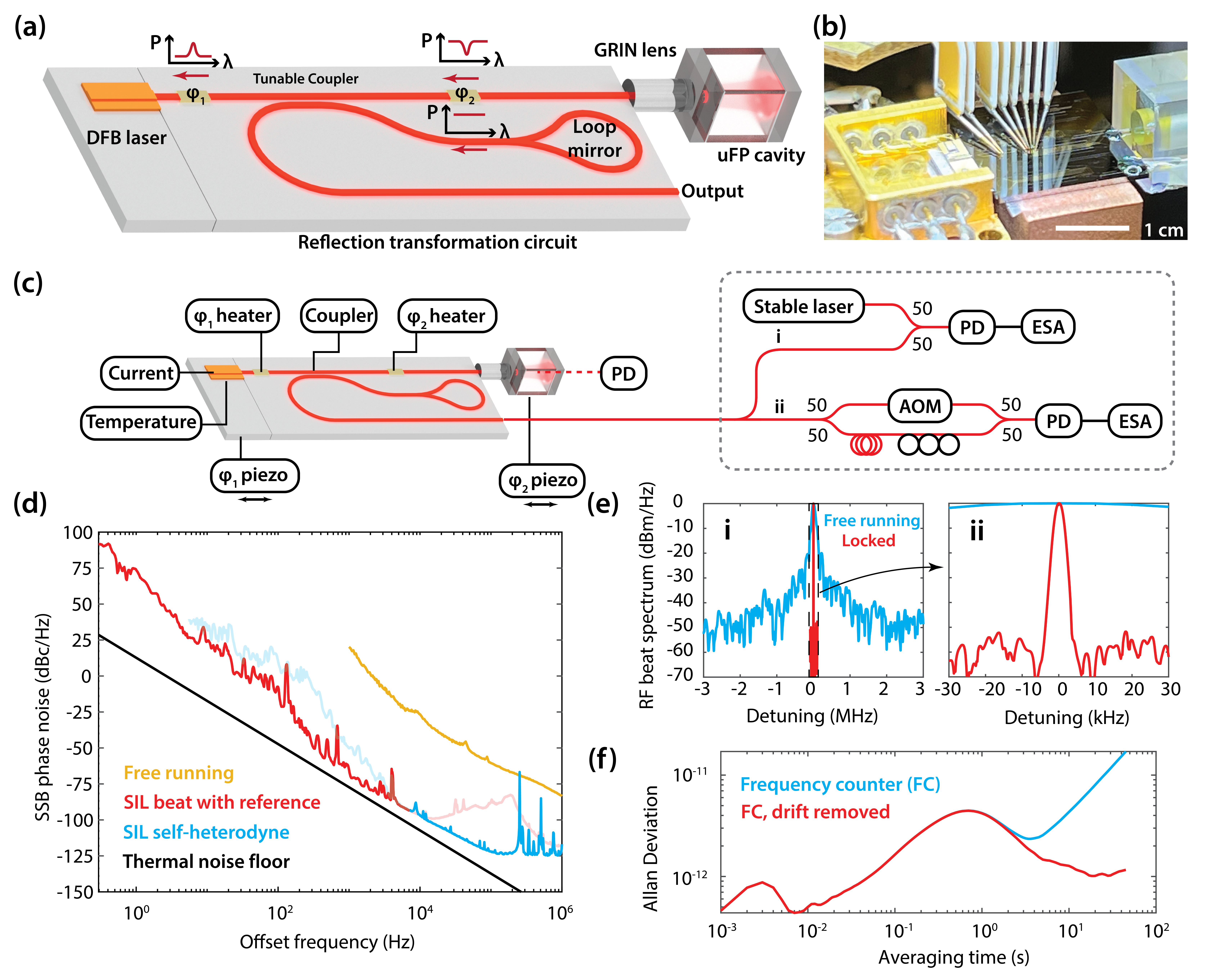}
    \caption{\label{fig2} \textbf{On-chip laser directly self-injection locked to co-integrated \textmu FP cavity. }
    \textbf{a}, Schematic illustration of on-chip circuit design enabling self-injection locking to co-integrated \textmu FP cavity.
    \textbf{b}, Photograph of DFB laser self-injection locked to \textmu FP cavity through interface chip.
    \textbf{c}, Schematic of experimental setup for noise measurement. Black lines indicates electrical control, and red lines indicates optical circuits. PD: photo detector. ESA: electrical spectrum analysizer. AOM: acousto-optical modulator. The output of SIL laser either beats with another stable laser or performs self-heterodyne with 800~m delay line for the ESA to measure phase noise. 
    \textbf{d}, Phase noise spectrum for SIL and free running laser. Free running laser phase noise is measured using self-heterodyne. The faded curve indicates the noise is limited by the measurement noise floor. 
    \textbf{e}, RF beat note spectrum of the SIL laser and a stable laser revels linewidth compression when locked. \textbf{ii} is the zoom in for \textbf{i}. Free running beat note is taken with 91~kHz RBW, and locked beat note is taken with 3~kHz RBW.
    \textbf{f}, Allan deviation of the SIL laser measured from frequency counter with and without linear drift removal. 
    }
\end{figure*}

\section{\label{sec:Results} Results}

\subsection{\label{subsec:End-fire} Photonic interface for self-injection locking}

Self-injection locking (SIL) utilizes optical feedback to lock a laser frequency's to an external cavity, causing the laser to inherit the noise characteristics of the cavity. 
Recent studies have used silicon nitride ring resonators to self-injection lock chip-scale semiconductor lasers, greatly improving their phase noise characteristics~\cite{jin2021hertz,lihachev2022platicon,guo2022chip,xiang20233d}.  
These demonstrations have relied on Rayleigh back-scattering to produce resonant feedback necessary for self-injection locking. 
However, thermo-refractive noise and scattering losses limit the achievable quality factors and noise characteristics of such waveguide-based resonators~\cite{liu2022ultralow,he2024chip}. Alternatively, if such dielectric ring resonators can be replaced by vacuum-gap \textmu FP reference cavities, thermorefractive noise could be practically eliminated, offering a path to lasers with greatly improved phase noise and frequency stability. 
However, self-injection locking of integrated lasers will require new strategies to interface such \textmu FP cavities with photonic integrated circuits to produce resonant feedback.

Established methods for producing resonant feedback have relied on circulators to redirect the transmission response of the FP cavity into the laser \cite{liang2023compact}. However, this strategy would be challenging to implement using photonic integrated circuits, as isolators and circulators are not widely available; moreover, it would be difficult to access both ports of the \textmu FP cavity using a single chip. 
Alternatively, resonant feedback can be produced by partial excitation of higher order modes using a tilted cavity configuration; however, this approach requires a more complex alignment procedure, and it results in excess coupling losses that unnecessarily reduce the efficiency of laser feedback \cite{dahmani1987frequency,laurent1989frequency,hjelme1991semiconductor,savchenkov2024robust}.

To address these challenges, we demonstrate a new circuit interface that transforms the reflection response of the \textmu FP cavity to produce resonant feedback necessary for self-injection locking.
In reflection, the cavity modes of an FP resonator appear as anti-resonances (dips) atop a high reflectivity background; such anti-resonances are unsuitable for self-injection locking.
These antiresonances result from interference between light that is resonantly scattered by the cavity modes and a broadband background reflection produced by the mirrors.
To utilize the \textmu FP for self-injection locking, it is crucial to eliminate the high reflectivity background, leaving only resonant scattering (or cavity leakage field) to produce feedback; this is because broadband mirror reflections can produce parasitic resonances that destabilize the laser~\cite{tkach1986regimes}. 
If the wideband mirror reflection can be eliminated, leaving only resonant backscatter from the cavity modes, efficient resonant feedback can be achieved.
In this case, the sudden phase shift produced by resonant backscattering about the cavity resonance provides the optimal feedback mechanism for self-injection locking.

The circuit interface that we use to produce resonant feedback into a waveguide-based DFB laser is seen in Fig.~2(a).
An on-chip directional coupler divides the incoming light, directing a portion of light into an on-chip loop-mirror that introduces an additional broadband mirror reflection. The remaining light travels to the chip facet, where it is collimated by a GRIN lens to match the mode of the co-integrated FP cavity. These two reflections interfere upon recombination at the directional coupler. With the correct phase $\phi_2$, the loop-mirror reflection destructively interferes with the FP cavity reflection, cancelling the broadband mirror reflections and preserving only the resonant backscattered light. This results in a resonant (Lorenzian) feedback spectrum, facilitating self-injection locking by adjusting the feedback phase $\phi_1$.

For the choice of GRIN lens, it is crucial to expand and collimate the beam of light emitted from the waveguide end-facet to match the \textmu FP cavity mode. To accomplish this mode transformation using our circuit interface, we use an inverse taper at the output facet of a $\mathrm{Si_3N_4}$ waveguide, producing a mode diameter of approximately 3~$\mathrm{\mu m}$. A GRIN lens with an effective focal length of 0.46~$\mathrm{mm}$ is then used to expand this 3~$\mathrm{\mu m}$ to 240~$\mathrm{\mu m}$, matching the mode field diameter of the \textmu FP cavity, permitting coupling losses as low as 1.5 dB. 
Hence, this simple and compact photonic interfacing technique, shown in Fig.~2(b), offers a practical means of co-integrating the \textmu FP cavity with PICs, to enable high performance oscillators and lasers for next-generation photonic systems.

\begin{figure}[!htb]
    \includegraphics{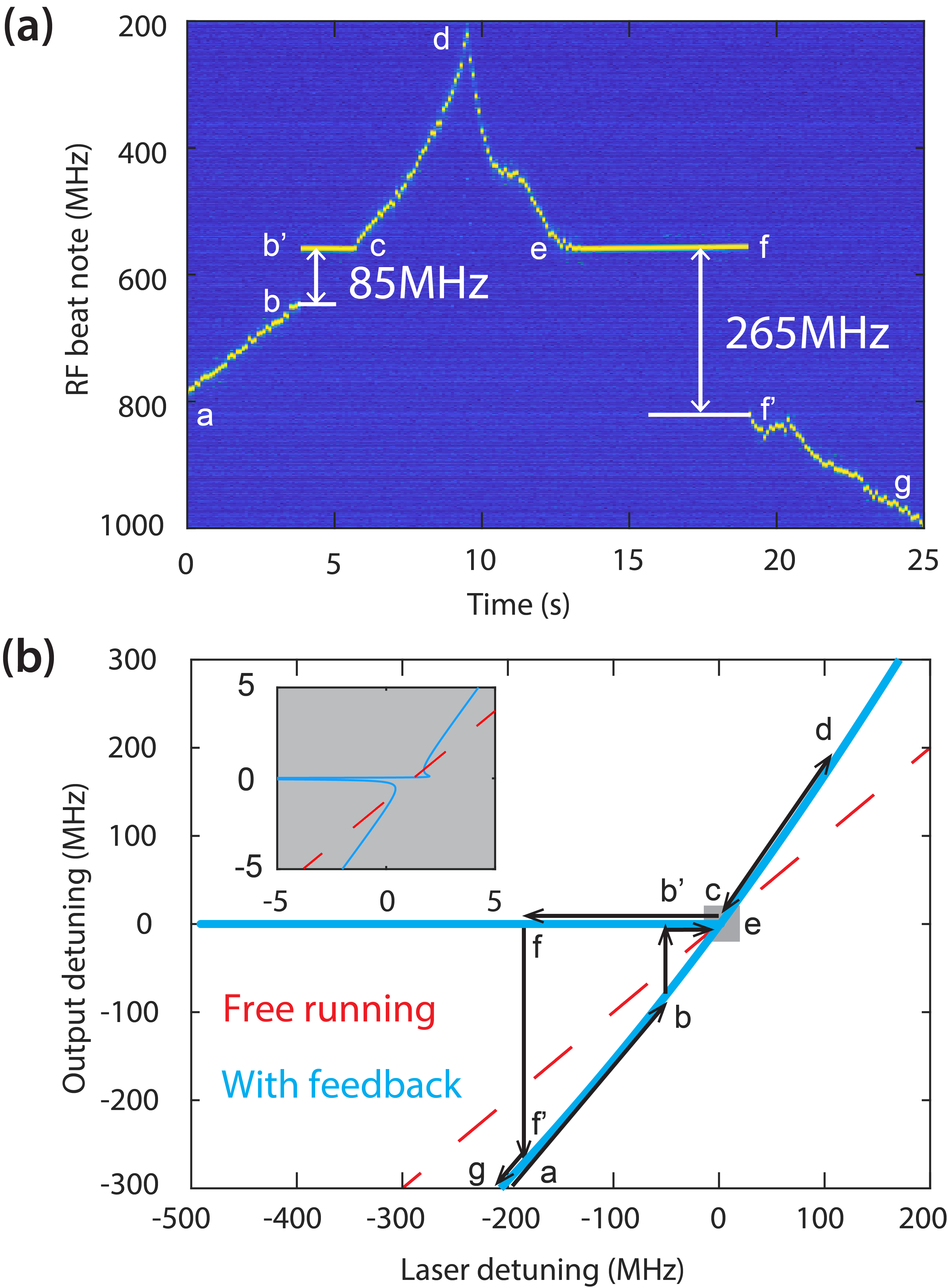}
    \centering
    \caption{\label{fig3} \textbf{Locking range and dynamics of the SIL laser. }
    \textbf{a}, RF beat note spectrum of the SIL laser and a stable laser with a sweeping current on the DFB laser yields a 265~MHz locking range.
    \textbf{b}, Theoretical locking curve. The SIL laser entering and exiting the locked state are due to relaxation oscillation of the DFB laser. Off-resonance curve deviates from free running due to the residual broadband mirror reflection.
    }
\end{figure}

The circuit interface of Fig.~2(c) is used to control the optical feedback parameters of this system, as necessary to obtain a stable self-injection locking. 
The interface circuit includes integrated heaters that allow for adjustment of the coupler splitting ratio, the feedback phase $\phi_1$, and the interferometer phase $\phi_2$.  
Piezo actuators on both the DFB laser and the \textmu FP cavity are also used for fine-tuning $\phi_1$ and $\phi_2$ by modulating their separation from the photonic circuit. 
When $\phi_1$ and $\phi_2$ are adjusted to produce a self-injection locked state, the frequency of the laser emission becomes resonant with the \textmu FP cavity, producing a dramatic increase in the light intensity within the \textmu FP cavity.
Light transmitted from the \textmu FP cavity is collected through free space and monitored with a photodetector to observe the onset of the self-injection locking process. 
The output from the SIL laser is then coupled into the fiber-based apparatus of Fig.~2(c,\textbf{i}) for phase noise measurements.

Two complementary techniques are used to quantify the phase noise of the self-injection-locked laser.
To measure the phase noise at low offset frequencies, we analyze the beat note between the integrated SIL laser and another frequency-stabilized reference laser using a phase noise analyzer, as shown in Fig.~2(c,\textbf{i}). The stable laser is comprised of a fiber laser that is PDH-locked to a 10~cm long table-top FP reference cavity housed in a vacuum chamber that is known to have low phase noise at low offset frequency (0~dBc/Hz at 1~Hz offset frequency).
However, since the PDH feedback loop introduces excess noise, seen as a servo bump at offset frequencies above 10~kHz, a complementary measurement method is required to analyze the phase noise of SIL laser at higher offset frequencies. 
Here we use the fiber interferometer with 800~m delay seen in Fig.~2(c,\textbf{ii}) to perform self-heterodyne measurements.
This complementary approach is ideal for phase noise measurements at high offset frequencies since the fiber interferometer is essentially free of technical noise for frequencies above 5 kHz.
Combining these two techniques, we obtain a complete picture of phase noise spectrum of the SIL laser from 1Hz to 1MHz, as seen in Fig.~2(d).  

The measurements of Fig.~2(d) reveal a dramatic reduction in the phase noise of the DFB laser when it is injection locked to the \textmu FP cavity.   
Self-injection locking is seen to reduce the phase noise of the DFB laser by eight, seven, and six orders of magnitude at offset frequencies of 1 kHz, 10 kHz, and 100 kHz, corresponding to phase noise levels of $-$65~dBc/Hz, $-$97~dBc/Hz, and $-$122~dBc/Hz, respectively.
Further analysis of the phase noise spectrum reveals a fundamental linewidth of 35 mHz and a 1/$\pi$ integral linewidth of 150 Hz. The reduction in laser linewidth produced by self-injection locking is further demonstrated in the direct RF beat note spectrum between the SIL laser and a stable reference laser, as shown in Fig.~2(e). It is important to note that the linewidth in Fig.~2(e,\textbf{ii}) is constrained by the resolution bandwidth (RBW) of the electronic spectrum analyzer. Note that these measurements were performed with a relatively large resolution bandwidth ($>$3 kHz) to ensure a stable beat note spectrum within the acquisition time.  
The Allan deviation, which is indicative of fractional frequency stability for different averaging times, is seen in Fig.~2(f), revealing $5\times10^{-13}$ at 10~ms averaging time. After removing linear drift of 109~Hz/s, the system exhibits a fractional frequency stability better than $5\times 10^{-12}$ from 1 ms to 50 s averaging times. 

Note that these measurements correspond to record-level phase noise among state-of-the-art chip-integrated systems\cite{he2024chip} and were obtained without tight environmental controls. 
The measurements described above were obtained without active temperature stabilization of the cavity, illustrating the temperature insensitivity and robustness of such in-vacuum bonded \textmu FP reference cavities. 
Deviations from the thermal noise limit at low offset frequencies are attributed to mechanical noise from the translation stages used to position the cavity and DFB laser. 
Hence, the low offset phase noise and frequency drift are likely to greatly improve when this system is fully co-integrated.

We also analyze the dynamics of the self-injection locking by slowly modulating the DFB laser current to observe shifts in the RF beat note spectrum with a stable reference laser, as depicted in Fig.~3(a). 
This system enters and exits the SIL state at points b to b' and f to f', respectively, demonstrating a locking range of 265 MHz. Comparison with a theoretical model of the self-injection locking process suggests that the onset of self-injection locking occurs earlier than expected; this is likely due to laser relaxation oscillations produced by the semiconductor gain medium~\cite{xiang20233d}. The asymmetric feature for two directions of SIL is due to the fact that the feedback phase $\phi_1$ is non-zero~\cite{xiang20233d}. It is also notable that away from the \textmu FP resonance, the relationship between output frequency and DFB laser frequency diverges from the identity line, indicating that mirror reflections are not entirely cancelled, leaving a small amount of residual broadband feedback. Nevertheless, provided that this feedback remains sufficiently weak to avoid destabilizing the laser, it appears to have little impact on the performance of the SIL laser.
\begin{figure*}[!htb]
    \includegraphics{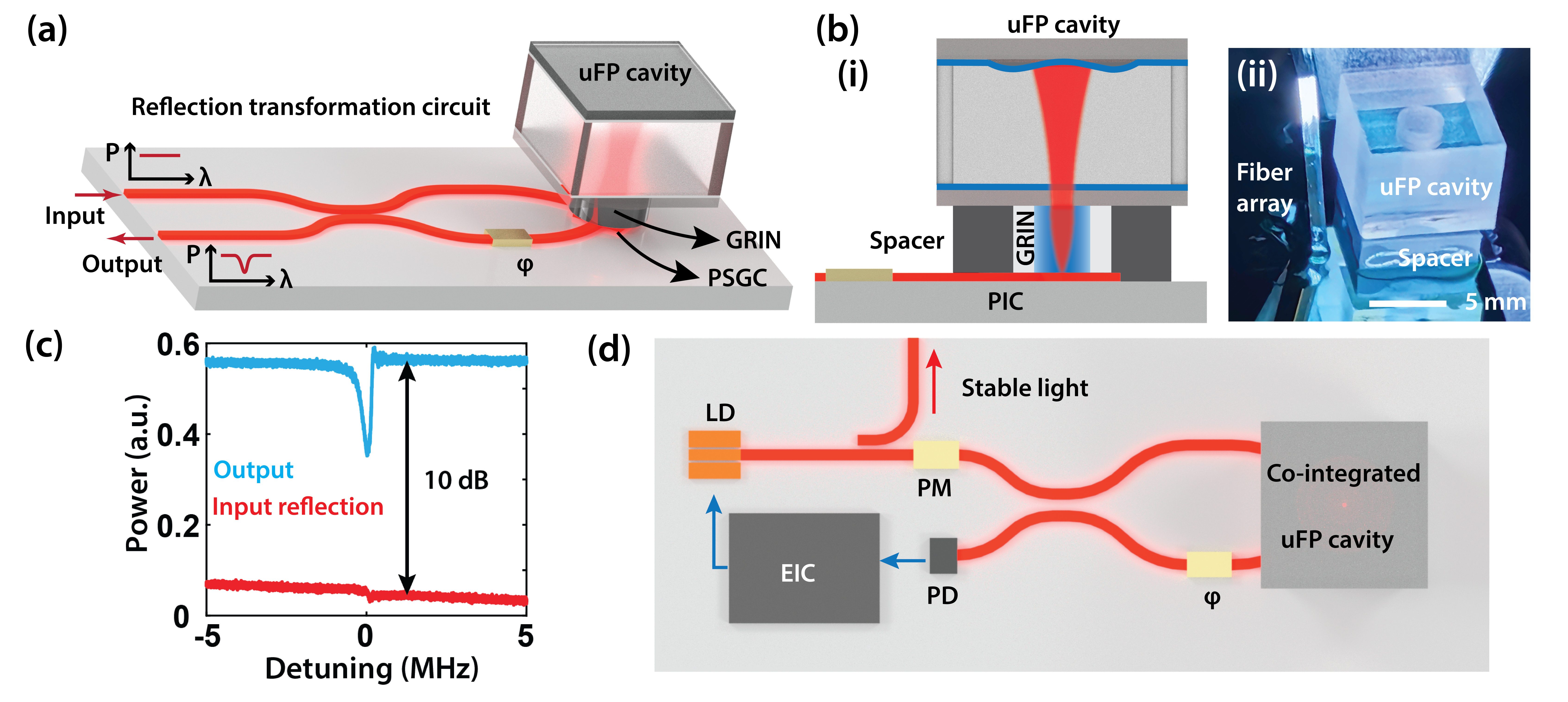}
    \centering
    \caption{\label{fig4} \textbf{co-integrated \textmu FP for electrical feedback.}
    \textbf{a}, Schematic of the approach for co-integrating \textmu FP cavity with signal redirection and back-reflection isolation.
    \textbf{b}, \textbf{i} Cross section of co-integrated \textmu FP cavity. \textbf{ii} Optical picture for co-integrated module.
    \textbf{c}, Experimental data validating signal redirection and characterizing back-reflection isolation ratio.
    \textbf{d}, Outlook for on-chip PDH locking system. PM: phase modulator. PD: photo detector. LD: laser diode. EIC: electric integrated circuits.
    }
\end{figure*}

\subsection{\label{subsec:Grating coupler} Photonic interface for electrical control}

Beyond self-injection locking, new circuit interfaces for efficient interrogation of \textmu FP cavity modes can enable complementary applications involving on-chip frequency metrology and microwave photonic signal processing.
For example, PDH-locking provides a very versitile means of harnessing reference cavities with many practical advantages. 
Since PDH-locking can be implemented with low incident optical powers ($\sim\! 10$~\textmu W) relative to self-injection locking ($>\! 1$~mW), photothermal noise is greatly reduced, permitting us to more readily reach the thermal noise limit (Fig. 2d). 
Additionally, PDH-locking enables frequency tunability and lower phase noise at low offset frequencies, making it indispensable for the most demanding applications requiring precise frequency control \cite{liu2024ultrastable, kudelin2024photonic}. 
Alternatively, such ultra-high Q-factor \textmu FP cavities can be used for microwave-photonic filtering and signal processing~\cite{marpaung2013si}. 
Applications such as PDH-locking and microwave photonic signal processing require a photonic interface for efficient interrogation of \textmu FP cavities while protecting on-chip lasers from strong feedback and reflections. Optical circulators are often used for redirecting reflected light for such applications~\cite{herrmann2022mirror,cheng2024terahertz}, but circulators and isolators are not yet widely available in integrated photonics due to the incompatibility of magnetic media with CMOS fabrication processes.

To address this challenge, we use a novel reflection cancellation circuit that adapts conventional schemes for a poor man’s isolator into a circuit architecture~\cite{cheng2023novel}.
This second strategy for interfacing and integrating the \textmu FP cavity uses a silicon photonic circuit fabricated from a silicon-on-insulator platform.
This circuit is comprised of a two-port interferometer that connects to an inverse-designed two-port polarization-splitting grating coupler (PSGC), as described in~\cite{cheng2023novel}.
By fine-tuning the on-chip phase shifter, this vertical-emission PSGC efficiently couples circularly polarized light into the \textmu FP cavity, permitting interrogation of the reflection response of the cavity. 
Making use of the fact that the chirality of circularly polarized light is reversed upon the reflection from \textmu FP cavity, this circuit achieves a function very similar to that of a circulator~\cite{cheng2023novel}.

Efficient coupling into the \textmu FP is achieved by using a GRIN lens to expand the beam produced by PSGC. 
The PSGC is designed to produce vertical emission of a 10.4~\textmu m mode field diameter.
This beam is expanded to produce a collimated 240~\textmu m mode field diameter at the input \textmu FP cavity using a GRIN lens with an effective focal length of 0.94~mm.
The GRIN lens and \textmu FP cavity are then co-integrated with PIC using a custom spacer that provides mechanical stability for the bonded assembly, as is shown in Fig~4(b).
Using this integration method, light can be coupled from the silicon circuit into the \textmu FP cavity with an efficiency of -2.8 dB.  
A fiber array was bonded the chip-integrated \textmu FP module to interrogate the response of this multi-port photonic interface.

Figure 4(c) shows the optical reflection from Port A and the transmission to Port B as a function of laser detuning from the cavity resonance.
As seen in Fig.~4(c), this circuit maps the \textmu FP cavity reflection response to output port while simultaneously suppressing back-reflections from input port.
The relative magnitude of the transmission and reflection reveal 10 dB back-reflection suppression, to aid in protecting an on-chip laser from unwanted feedback.
Such a fully integrated module is readily suitable for an on-chip PDH locking system (Fig.~4(d)) as proposed in \cite{kudelin2024photonic}. Together with many other applications such as photonic RF filtering and microwave signal generators, the co-integration of vacuum-gap \textmu FP cavities paves the way for a next-generation of compact, high-performance RF photonic systems.

\section{\label{sec:disc} Discussion and conclusion}

We have demonstrated complementary approaches for the integration of high-performance vacuum-gap \textmu FP reference cavities with photonic integrated circuits, to enable both electrical and optical feedback for next-generation lasers, oscillators, and sensors.
Using a new circuit interface to transform the response of the \textmu FP resonator, we have created resonant optical feedback necessary to self-injection lock a waveguide-based laser with the reference cavity. This strategy produces an integrated laser with a fundamental linewidth of 35~mHz and an integrated 1/$\pi$ linewidth of 150 Hz, corresponding to record-level performance. 
Phase noise measurements represent a 19 dB (17 dB) improvement in phase noise at 1 kHz (10 kHz) offset frequencies relative to prior studies that have utilized 1.4~m spiral waveguide resonators\cite{he2024chip}.
Since the electromagnetic energy is stored almost entirely in vacuum, this vacuum-gap  \textmu FP cavity overcomes numerous limitations imposed by dielectric-induced thermo-refractive noise. Moving from stage controlled pieces to a fully packaged and on-chip electric-optical-controlled system, we anticipate the low-offset frequency noise will be further improved and long-term locking robustness will be extended from several hours to longer durations.
Such low phase noise levels make this system highly suitable for photonic microwave signal generators through use of optical frequency division, potentially eliminating the need for more complex PDH locking schemes\cite{kudelin2024photonic}.

Through a complementary approach, we have shown that the reflection response of such vacuum-gap uFP cavities can be interrogated using a circuit interface, to enable on-chip PDH-locking and microwave photonic signal processing. 
Co-integration of the cavity with a silicon photonic circuit is achieved by attaching the cavity to a multi-port vertical-emission grating coupler.
Using this grating interface, the cavity is placed within an optical interferometer that cancels unwanted back-reflections, eliminating the need for on-chip circulators and isolators.
This same interferometer maps the reflection response to a separate optical port, permitting efficient interrogation of the reflection response required for PDH-locking and microwave photonic signal processing.
Hence, this complementary interfacing method provides access to an ultra-narrow linewidth and stable reference cavity that becomes a versatile resource for advanced microwave-photonic filtering, on-chip PDH locking, and sensing applications.

Looking beyond these specific demonstrations, the co-integration of such high-performance vacuum-gap reference cavities with photonic circuits presents numerous advantages. By storing electro-magnetic field mostly in vacuum, it mitigates the limitations associated with thermo-refractive noise and enables high performance across various photonic integrated systems. Additionally, with sealed vacuum and athermal material, these cavities significantly reduce environmental constraints, thereby broadening the operating conditions. Furthermore, such reference cavities can achieve substantially higher performance with larger mode volumes, increased cavity lengths, and improved packaging. Hence, these innovative techniques provide a crucial capabilities that are essential for the development of next-generation integrated lasers, oscillators, and sensors.

\bibliography{MainRef}

\begin{thebibliography}{33}%
\makeatletter
\providecommand \@ifxundefined [1]{%
 \@ifx{#1\undefined}
}%
\providecommand \@ifnum [1]{%
 \ifnum #1\expandafter \@firstoftwo
 \else \expandafter \@secondoftwo
 \fi
}%
\providecommand \@ifx [1]{%
 \ifx #1\expandafter \@firstoftwo
 \else \expandafter \@secondoftwo
 \fi
}%
\providecommand \natexlab [1]{#1}%
\providecommand \enquote  [1]{``#1''}%
\providecommand \bibnamefont  [1]{#1}%
\providecommand \bibfnamefont [1]{#1}%
\providecommand \citenamefont [1]{#1}%
\providecommand \href@noop [0]{\@secondoftwo}%
\providecommand \href [0]{\begingroup \@sanitize@url \@href}%
\providecommand \@href[1]{\@@startlink{#1}\@@href}%
\providecommand \@@href[1]{\endgroup#1\@@endlink}%
\providecommand \@sanitize@url [0]{\catcode `\\12\catcode `\$12\catcode `\&12\catcode `\#12\catcode `\^12\catcode `\_12\catcode `\%12\relax}%
\providecommand \@@startlink[1]{}%
\providecommand \@@endlink[0]{}%
\providecommand \url  [0]{\begingroup\@sanitize@url \@url }%
\providecommand \@url [1]{\endgroup\@href {#1}{\urlprefix }}%
\providecommand \urlprefix  [0]{URL }%
\providecommand \Eprint [0]{\href }%
\providecommand \doibase [0]{https://doi.org/}%
\providecommand \selectlanguage [0]{\@gobble}%
\providecommand \bibinfo  [0]{\@secondoftwo}%
\providecommand \bibfield  [0]{\@secondoftwo}%
\providecommand \translation [1]{[#1]}%
\providecommand \BibitemOpen [0]{}%
\providecommand \bibitemStop [0]{}%
\providecommand \bibitemNoStop [0]{.\EOS\space}%
\providecommand \EOS [0]{\spacefactor3000\relax}%
\providecommand \BibitemShut  [1]{\csname bibitem#1\endcsname}%
\let\auto@bib@innerbib\@empty
\bibitem [{\citenamefont {Kessler}\ \emph {et~al.}(2012)\citenamefont {Kessler}, \citenamefont {Hagemann}, \citenamefont {Grebing}, \citenamefont {Legero}, \citenamefont {Sterr}, \citenamefont {Riehle}, \citenamefont {Martin}, \citenamefont {Chen},\ and\ \citenamefont {Ye}}]{kessler2012sub}%
  \BibitemOpen
  \bibfield  {author} {\bibinfo {author} {\bibfnamefont {T.}~\bibnamefont {Kessler}}, \bibinfo {author} {\bibfnamefont {C.}~\bibnamefont {Hagemann}}, \bibinfo {author} {\bibfnamefont {C.}~\bibnamefont {Grebing}}, \bibinfo {author} {\bibfnamefont {T.}~\bibnamefont {Legero}}, \bibinfo {author} {\bibfnamefont {U.}~\bibnamefont {Sterr}}, \bibinfo {author} {\bibfnamefont {F.}~\bibnamefont {Riehle}}, \bibinfo {author} {\bibfnamefont {M.}~\bibnamefont {Martin}}, \bibinfo {author} {\bibfnamefont {L.}~\bibnamefont {Chen}},\ and\ \bibinfo {author} {\bibfnamefont {J.}~\bibnamefont {Ye}},\ }\bibfield  {title} {\bibinfo {title} {A sub-40-m{H}z-linewidth laser based on a silicon single-crystal optical cavity},\ }\href@noop {} {\bibfield  {journal} {\bibinfo  {journal} {Nature Photonics}\ }\textbf {\bibinfo {volume} {6}},\ \bibinfo {pages} {687} (\bibinfo {year} {2012})}\BibitemShut {NoStop}%
\bibitem [{\citenamefont {Guo}\ \emph {et~al.}(2022)\citenamefont {Guo}, \citenamefont {McLemore}, \citenamefont {Xiang}, \citenamefont {Lee}, \citenamefont {Wu}, \citenamefont {Jin}, \citenamefont {Kelleher}, \citenamefont {Jin}, \citenamefont {Mason}, \citenamefont {Chang} \emph {et~al.}}]{guo2022chip}%
  \BibitemOpen
  \bibfield  {author} {\bibinfo {author} {\bibfnamefont {J.}~\bibnamefont {Guo}}, \bibinfo {author} {\bibfnamefont {C.~A.}\ \bibnamefont {McLemore}}, \bibinfo {author} {\bibfnamefont {C.}~\bibnamefont {Xiang}}, \bibinfo {author} {\bibfnamefont {D.}~\bibnamefont {Lee}}, \bibinfo {author} {\bibfnamefont {L.}~\bibnamefont {Wu}}, \bibinfo {author} {\bibfnamefont {W.}~\bibnamefont {Jin}}, \bibinfo {author} {\bibfnamefont {M.}~\bibnamefont {Kelleher}}, \bibinfo {author} {\bibfnamefont {N.}~\bibnamefont {Jin}}, \bibinfo {author} {\bibfnamefont {D.}~\bibnamefont {Mason}}, \bibinfo {author} {\bibfnamefont {L.}~\bibnamefont {Chang}}, \emph {et~al.},\ }\bibfield  {title} {\bibinfo {title} {Chip-based laser with 1-hertz integrated linewidth},\ }\href@noop {} {\bibfield  {journal} {\bibinfo  {journal} {Science advances}\ }\textbf {\bibinfo {volume} {8}},\ \bibinfo {pages} {eabp9006} (\bibinfo {year} {2022})}\BibitemShut {NoStop}%
\bibitem [{\citenamefont {Ludlow}\ \emph {et~al.}(2015)\citenamefont {Ludlow}, \citenamefont {Boyd}, \citenamefont {Ye}, \citenamefont {Peik},\ and\ \citenamefont {Schmidt}}]{ludlow2015optical}%
  \BibitemOpen
  \bibfield  {author} {\bibinfo {author} {\bibfnamefont {A.~D.}\ \bibnamefont {Ludlow}}, \bibinfo {author} {\bibfnamefont {M.~M.}\ \bibnamefont {Boyd}}, \bibinfo {author} {\bibfnamefont {J.}~\bibnamefont {Ye}}, \bibinfo {author} {\bibfnamefont {E.}~\bibnamefont {Peik}},\ and\ \bibinfo {author} {\bibfnamefont {P.~O.}\ \bibnamefont {Schmidt}},\ }\bibfield  {title} {\bibinfo {title} {Optical atomic clocks},\ }\href@noop {} {\bibfield  {journal} {\bibinfo  {journal} {Reviews of Modern Physics}\ }\textbf {\bibinfo {volume} {87}},\ \bibinfo {pages} {637} (\bibinfo {year} {2015})}\BibitemShut {NoStop}%
\bibitem [{\citenamefont {Liu}\ \emph {et~al.}(2024{\natexlab{a}})\citenamefont {Liu}, \citenamefont {Lee}, \citenamefont {Nakamura}, \citenamefont {Jin}, \citenamefont {Cheng}, \citenamefont {Kelleher}, \citenamefont {McLemore}, \citenamefont {Kudelin}, \citenamefont {Groman}, \citenamefont {Diddams} \emph {et~al.}}]{liu2024low}%
  \BibitemOpen
  \bibfield  {author} {\bibinfo {author} {\bibfnamefont {Y.}~\bibnamefont {Liu}}, \bibinfo {author} {\bibfnamefont {D.}~\bibnamefont {Lee}}, \bibinfo {author} {\bibfnamefont {T.}~\bibnamefont {Nakamura}}, \bibinfo {author} {\bibfnamefont {N.}~\bibnamefont {Jin}}, \bibinfo {author} {\bibfnamefont {H.}~\bibnamefont {Cheng}}, \bibinfo {author} {\bibfnamefont {M.~L.}\ \bibnamefont {Kelleher}}, \bibinfo {author} {\bibfnamefont {C.~A.}\ \bibnamefont {McLemore}}, \bibinfo {author} {\bibfnamefont {I.}~\bibnamefont {Kudelin}}, \bibinfo {author} {\bibfnamefont {W.}~\bibnamefont {Groman}}, \bibinfo {author} {\bibfnamefont {S.~A.}\ \bibnamefont {Diddams}}, \emph {et~al.},\ }\bibfield  {title} {\bibinfo {title} {Low-noise microwave generation with an air-gap optical reference cavity},\ }\href@noop {} {\bibfield  {journal} {\bibinfo  {journal} {APL Photonics}\ }\textbf {\bibinfo {volume} {9}} (\bibinfo {year} {2024}{\natexlab{a}})}\BibitemShut {NoStop}%
\bibitem [{\citenamefont {Kudelin}\ \emph {et~al.}(2024)\citenamefont {Kudelin}, \citenamefont {Groman}, \citenamefont {Ji}, \citenamefont {Guo}, \citenamefont {Kelleher}, \citenamefont {Lee}, \citenamefont {Nakamura}, \citenamefont {McLemore}, \citenamefont {Shirmohammadi}, \citenamefont {Hanifi} \emph {et~al.}}]{kudelin2024photonic}%
  \BibitemOpen
  \bibfield  {author} {\bibinfo {author} {\bibfnamefont {I.}~\bibnamefont {Kudelin}}, \bibinfo {author} {\bibfnamefont {W.}~\bibnamefont {Groman}}, \bibinfo {author} {\bibfnamefont {Q.-X.}\ \bibnamefont {Ji}}, \bibinfo {author} {\bibfnamefont {J.}~\bibnamefont {Guo}}, \bibinfo {author} {\bibfnamefont {M.~L.}\ \bibnamefont {Kelleher}}, \bibinfo {author} {\bibfnamefont {D.}~\bibnamefont {Lee}}, \bibinfo {author} {\bibfnamefont {T.}~\bibnamefont {Nakamura}}, \bibinfo {author} {\bibfnamefont {C.~A.}\ \bibnamefont {McLemore}}, \bibinfo {author} {\bibfnamefont {P.}~\bibnamefont {Shirmohammadi}}, \bibinfo {author} {\bibfnamefont {S.}~\bibnamefont {Hanifi}}, \emph {et~al.},\ }\bibfield  {title} {\bibinfo {title} {Photonic chip-based low-noise microwave oscillator},\ }\href@noop {} {\bibfield  {journal} {\bibinfo  {journal} {Nature}\ ,\ \bibinfo {pages} {1}} (\bibinfo {year} {2024})}\BibitemShut {NoStop}%
\bibitem [{\citenamefont {Kuhn}\ \emph {et~al.}(2002)\citenamefont {Kuhn}, \citenamefont {Hennrich},\ and\ \citenamefont {Rempe}}]{kuhn2002deterministic}%
  \BibitemOpen
  \bibfield  {author} {\bibinfo {author} {\bibfnamefont {A.}~\bibnamefont {Kuhn}}, \bibinfo {author} {\bibfnamefont {M.}~\bibnamefont {Hennrich}},\ and\ \bibinfo {author} {\bibfnamefont {G.}~\bibnamefont {Rempe}},\ }\bibfield  {title} {\bibinfo {title} {Deterministic single-photon source for distributed quantum networking},\ }\href@noop {} {\bibfield  {journal} {\bibinfo  {journal} {Physical review letters}\ }\textbf {\bibinfo {volume} {89}},\ \bibinfo {pages} {067901} (\bibinfo {year} {2002})}\BibitemShut {NoStop}%
\bibitem [{\citenamefont {Liang}\ \emph {et~al.}(2010)\citenamefont {Liang}, \citenamefont {Ilchenko}, \citenamefont {Savchenkov}, \citenamefont {Matsko}, \citenamefont {Seidel},\ and\ \citenamefont {Maleki}}]{liang2010whispering}%
  \BibitemOpen
  \bibfield  {author} {\bibinfo {author} {\bibfnamefont {W.}~\bibnamefont {Liang}}, \bibinfo {author} {\bibfnamefont {V.}~\bibnamefont {Ilchenko}}, \bibinfo {author} {\bibfnamefont {A.}~\bibnamefont {Savchenkov}}, \bibinfo {author} {\bibfnamefont {A.}~\bibnamefont {Matsko}}, \bibinfo {author} {\bibfnamefont {D.}~\bibnamefont {Seidel}},\ and\ \bibinfo {author} {\bibfnamefont {L.}~\bibnamefont {Maleki}},\ }\bibfield  {title} {\bibinfo {title} {Whispering-gallery-mode-resonator-based ultranarrow linewidth external-cavity semiconductor laser},\ }\href@noop {} {\bibfield  {journal} {\bibinfo  {journal} {Optics letters}\ }\textbf {\bibinfo {volume} {35}},\ \bibinfo {pages} {2822} (\bibinfo {year} {2010})}\BibitemShut {NoStop}%
\bibitem [{\citenamefont {Jin}\ \emph {et~al.}(2022{\natexlab{a}})\citenamefont {Jin}, \citenamefont {Mason}, \citenamefont {Luo}, \citenamefont {Kharel}, \citenamefont {Isichenko}, \citenamefont {Blumenthal}, \citenamefont {Papp},\ and\ \citenamefont {Rakich}}]{jin2022microfabrication}%
  \BibitemOpen
  \bibfield  {author} {\bibinfo {author} {\bibfnamefont {N.}~\bibnamefont {Jin}}, \bibinfo {author} {\bibfnamefont {D.}~\bibnamefont {Mason}}, \bibinfo {author} {\bibfnamefont {Y.}~\bibnamefont {Luo}}, \bibinfo {author} {\bibfnamefont {P.}~\bibnamefont {Kharel}}, \bibinfo {author} {\bibfnamefont {A.}~\bibnamefont {Isichenko}}, \bibinfo {author} {\bibfnamefont {D.~J.}\ \bibnamefont {Blumenthal}}, \bibinfo {author} {\bibfnamefont {S.~B.}\ \bibnamefont {Papp}},\ and\ \bibinfo {author} {\bibfnamefont {P.~T.}\ \bibnamefont {Rakich}},\ }\bibfield  {title} {\bibinfo {title} {Microfabrication of dielectric resonators with {Q}-factor exceeding 5 billion},\ }in\ \href@noop {} {\emph {\bibinfo {booktitle} {Frontiers in Optics}}}\ (\bibinfo {organization} {Optica Publishing Group},\ \bibinfo {year} {2022})\ pp.\ \bibinfo {pages} {FM4C--1}\BibitemShut {NoStop}%
\bibitem [{\citenamefont {Jin}\ \emph {et~al.}(2022{\natexlab{b}})\citenamefont {Jin}, \citenamefont {McLemore}, \citenamefont {Mason}, \citenamefont {Hendrie}, \citenamefont {Luo}, \citenamefont {Kelleher}, \citenamefont {Kharel}, \citenamefont {Quinlan}, \citenamefont {Diddams},\ and\ \citenamefont {Rakich}}]{jin2022micro}%
  \BibitemOpen
  \bibfield  {author} {\bibinfo {author} {\bibfnamefont {N.}~\bibnamefont {Jin}}, \bibinfo {author} {\bibfnamefont {C.~A.}\ \bibnamefont {McLemore}}, \bibinfo {author} {\bibfnamefont {D.}~\bibnamefont {Mason}}, \bibinfo {author} {\bibfnamefont {J.~P.}\ \bibnamefont {Hendrie}}, \bibinfo {author} {\bibfnamefont {Y.}~\bibnamefont {Luo}}, \bibinfo {author} {\bibfnamefont {M.~L.}\ \bibnamefont {Kelleher}}, \bibinfo {author} {\bibfnamefont {P.}~\bibnamefont {Kharel}}, \bibinfo {author} {\bibfnamefont {F.}~\bibnamefont {Quinlan}}, \bibinfo {author} {\bibfnamefont {S.~A.}\ \bibnamefont {Diddams}},\ and\ \bibinfo {author} {\bibfnamefont {P.~T.}\ \bibnamefont {Rakich}},\ }\bibfield  {title} {\bibinfo {title} {Micro-fabricated mirrors with finesse exceeding one million},\ }\href@noop {} {\bibfield  {journal} {\bibinfo  {journal} {Optica}\ }\textbf {\bibinfo {volume} {9}},\ \bibinfo {pages} {965} (\bibinfo {year} {2022}{\natexlab{b}})}\BibitemShut {NoStop}%
\bibitem [{\citenamefont {Liu}\ \emph {et~al.}(2024{\natexlab{b}})\citenamefont {Liu}, \citenamefont {Jin}, \citenamefont {Lee}, \citenamefont {McLemore}, \citenamefont {Nakamura}, \citenamefont {Kelleher}, \citenamefont {Cheng}, \citenamefont {Schima}, \citenamefont {Hoghooghi}, \citenamefont {Diddams} \emph {et~al.}}]{liu2024ultrastable}%
  \BibitemOpen
  \bibfield  {author} {\bibinfo {author} {\bibfnamefont {Y.}~\bibnamefont {Liu}}, \bibinfo {author} {\bibfnamefont {N.}~\bibnamefont {Jin}}, \bibinfo {author} {\bibfnamefont {D.}~\bibnamefont {Lee}}, \bibinfo {author} {\bibfnamefont {C.}~\bibnamefont {McLemore}}, \bibinfo {author} {\bibfnamefont {T.}~\bibnamefont {Nakamura}}, \bibinfo {author} {\bibfnamefont {M.}~\bibnamefont {Kelleher}}, \bibinfo {author} {\bibfnamefont {H.}~\bibnamefont {Cheng}}, \bibinfo {author} {\bibfnamefont {S.}~\bibnamefont {Schima}}, \bibinfo {author} {\bibfnamefont {N.}~\bibnamefont {Hoghooghi}}, \bibinfo {author} {\bibfnamefont {S.}~\bibnamefont {Diddams}}, \emph {et~al.},\ }\bibfield  {title} {\bibinfo {title} {Ultrastable vacuum-gap {F}abry-{P}erot cavities operated in air},\ }\href@noop {} {\bibfield  {journal} {\bibinfo  {journal} {arXiv preprint arXiv:2406.13159}\ } (\bibinfo {year} {2024}{\natexlab{b}})}\BibitemShut {NoStop}%
\bibitem [{\citenamefont {Cheng}\ \emph {et~al.}(2023)\citenamefont {Cheng}, \citenamefont {Jin}, \citenamefont {Dai}, \citenamefont {Xiang}, \citenamefont {Guo}, \citenamefont {Zhou}, \citenamefont {Diddams}, \citenamefont {Quinlan}, \citenamefont {Bowers}, \citenamefont {Miller} \emph {et~al.}}]{cheng2023novel}%
  \BibitemOpen
  \bibfield  {author} {\bibinfo {author} {\bibfnamefont {H.}~\bibnamefont {Cheng}}, \bibinfo {author} {\bibfnamefont {N.}~\bibnamefont {Jin}}, \bibinfo {author} {\bibfnamefont {Z.}~\bibnamefont {Dai}}, \bibinfo {author} {\bibfnamefont {C.}~\bibnamefont {Xiang}}, \bibinfo {author} {\bibfnamefont {J.}~\bibnamefont {Guo}}, \bibinfo {author} {\bibfnamefont {Y.}~\bibnamefont {Zhou}}, \bibinfo {author} {\bibfnamefont {S.~A.}\ \bibnamefont {Diddams}}, \bibinfo {author} {\bibfnamefont {F.}~\bibnamefont {Quinlan}}, \bibinfo {author} {\bibfnamefont {J.}~\bibnamefont {Bowers}}, \bibinfo {author} {\bibfnamefont {O.}~\bibnamefont {Miller}}, \emph {et~al.},\ }\bibfield  {title} {\bibinfo {title} {A novel approach to interface high-{Q} {F}abry--{P}rot resonators with photonic circuits},\ }\href@noop {} {\bibfield  {journal} {\bibinfo  {journal} {APL Photonics}\ }\textbf {\bibinfo {volume} {8}} (\bibinfo {year} {2023})}\BibitemShut {NoStop}%
\bibitem [{\citenamefont {Kondratiev}\ \emph {et~al.}(2017)\citenamefont {Kondratiev}, \citenamefont {Lobanov}, \citenamefont {Cherenkov}, \citenamefont {Voloshin}, \citenamefont {Pavlov}, \citenamefont {Koptyaev},\ and\ \citenamefont {Gorodetsky}}]{kondratiev2017self}%
  \BibitemOpen
  \bibfield  {author} {\bibinfo {author} {\bibfnamefont {N.}~\bibnamefont {Kondratiev}}, \bibinfo {author} {\bibfnamefont {V.}~\bibnamefont {Lobanov}}, \bibinfo {author} {\bibfnamefont {A.}~\bibnamefont {Cherenkov}}, \bibinfo {author} {\bibfnamefont {A.}~\bibnamefont {Voloshin}}, \bibinfo {author} {\bibfnamefont {N.}~\bibnamefont {Pavlov}}, \bibinfo {author} {\bibfnamefont {S.}~\bibnamefont {Koptyaev}},\ and\ \bibinfo {author} {\bibfnamefont {M.}~\bibnamefont {Gorodetsky}},\ }\bibfield  {title} {\bibinfo {title} {Self-injection locking of a laser diode to a high-{Q} {WGM} microresonator},\ }\href@noop {} {\bibfield  {journal} {\bibinfo  {journal} {Optics Express}\ }\textbf {\bibinfo {volume} {25}},\ \bibinfo {pages} {28167} (\bibinfo {year} {2017})}\BibitemShut {NoStop}%
\bibitem [{\citenamefont {Tkach}\ and\ \citenamefont {Chraplyvy}(1986)}]{tkach1986regimes}%
  \BibitemOpen
  \bibfield  {author} {\bibinfo {author} {\bibfnamefont {R.}~\bibnamefont {Tkach}}\ and\ \bibinfo {author} {\bibfnamefont {A.}~\bibnamefont {Chraplyvy}},\ }\bibfield  {title} {\bibinfo {title} {Regimes of feedback effects in 1.5-$\mu$m distributed feedback lasers},\ }\href@noop {} {\bibfield  {journal} {\bibinfo  {journal} {Journal of Lightwave technology}\ }\textbf {\bibinfo {volume} {4}},\ \bibinfo {pages} {1655} (\bibinfo {year} {1986})}\BibitemShut {NoStop}%
\bibitem [{\citenamefont {Gundavarapu}\ \emph {et~al.}(2019)\citenamefont {Gundavarapu}, \citenamefont {Brodnik}, \citenamefont {Puckett}, \citenamefont {Huffman}, \citenamefont {Bose}, \citenamefont {Behunin}, \citenamefont {Wu}, \citenamefont {Qiu}, \citenamefont {Pinho}, \citenamefont {Chauhan} \emph {et~al.}}]{gundavarapu2019sub}%
  \BibitemOpen
  \bibfield  {author} {\bibinfo {author} {\bibfnamefont {S.}~\bibnamefont {Gundavarapu}}, \bibinfo {author} {\bibfnamefont {G.~M.}\ \bibnamefont {Brodnik}}, \bibinfo {author} {\bibfnamefont {M.}~\bibnamefont {Puckett}}, \bibinfo {author} {\bibfnamefont {T.}~\bibnamefont {Huffman}}, \bibinfo {author} {\bibfnamefont {D.}~\bibnamefont {Bose}}, \bibinfo {author} {\bibfnamefont {R.}~\bibnamefont {Behunin}}, \bibinfo {author} {\bibfnamefont {J.}~\bibnamefont {Wu}}, \bibinfo {author} {\bibfnamefont {T.}~\bibnamefont {Qiu}}, \bibinfo {author} {\bibfnamefont {C.}~\bibnamefont {Pinho}}, \bibinfo {author} {\bibfnamefont {N.}~\bibnamefont {Chauhan}}, \emph {et~al.},\ }\bibfield  {title} {\bibinfo {title} {Sub-hertz fundamental linewidth photonic integrated brillouin laser},\ }\href@noop {} {\bibfield  {journal} {\bibinfo  {journal} {Nature Photonics}\ }\textbf {\bibinfo {volume} {13}},\ \bibinfo {pages} {60} (\bibinfo {year} {2019})}\BibitemShut {NoStop}%
\bibitem [{\citenamefont {Lai}\ \emph {et~al.}(2022)\citenamefont {Lai}, \citenamefont {El~Amili}, \citenamefont {Eliyahu}, \citenamefont {Moss}, \citenamefont {Ganji}, \citenamefont {Singer},\ and\ \citenamefont {Maleki}}]{lai2022ultra}%
  \BibitemOpen
  \bibfield  {author} {\bibinfo {author} {\bibfnamefont {Y.-H.}\ \bibnamefont {Lai}}, \bibinfo {author} {\bibfnamefont {A.}~\bibnamefont {El~Amili}}, \bibinfo {author} {\bibfnamefont {D.}~\bibnamefont {Eliyahu}}, \bibinfo {author} {\bibfnamefont {R.}~\bibnamefont {Moss}}, \bibinfo {author} {\bibfnamefont {S.}~\bibnamefont {Ganji}}, \bibinfo {author} {\bibfnamefont {S.}~\bibnamefont {Singer}},\ and\ \bibinfo {author} {\bibfnamefont {L.}~\bibnamefont {Maleki}},\ }\bibfield  {title} {\bibinfo {title} {Ultra-narrow-linewidth lasers for quantum applications},\ }in\ \href@noop {} {\emph {\bibinfo {booktitle} {2022 Conference on Lasers and Electro-Optics (CLEO)}}}\ (\bibinfo {organization} {IEEE},\ \bibinfo {year} {2022})\ pp.\ \bibinfo {pages} {1--2}\BibitemShut {NoStop}%
\bibitem [{\citenamefont {Jin}\ \emph {et~al.}(2021)\citenamefont {Jin}, \citenamefont {Yang}, \citenamefont {Chang}, \citenamefont {Shen}, \citenamefont {Wang}, \citenamefont {Leal}, \citenamefont {Wu}, \citenamefont {Gao}, \citenamefont {Feshali}, \citenamefont {Paniccia} \emph {et~al.}}]{jin2021hertz}%
  \BibitemOpen
  \bibfield  {author} {\bibinfo {author} {\bibfnamefont {W.}~\bibnamefont {Jin}}, \bibinfo {author} {\bibfnamefont {Q.-F.}\ \bibnamefont {Yang}}, \bibinfo {author} {\bibfnamefont {L.}~\bibnamefont {Chang}}, \bibinfo {author} {\bibfnamefont {B.}~\bibnamefont {Shen}}, \bibinfo {author} {\bibfnamefont {H.}~\bibnamefont {Wang}}, \bibinfo {author} {\bibfnamefont {M.~A.}\ \bibnamefont {Leal}}, \bibinfo {author} {\bibfnamefont {L.}~\bibnamefont {Wu}}, \bibinfo {author} {\bibfnamefont {M.}~\bibnamefont {Gao}}, \bibinfo {author} {\bibfnamefont {A.}~\bibnamefont {Feshali}}, \bibinfo {author} {\bibfnamefont {M.}~\bibnamefont {Paniccia}}, \emph {et~al.},\ }\bibfield  {title} {\bibinfo {title} {Hertz-linewidth semiconductor lasers using {CMOS}-ready ultra-high-{Q} microresonators},\ }\href@noop {} {\bibfield  {journal} {\bibinfo  {journal} {Nature Photonics}\ }\textbf {\bibinfo {volume} {15}},\ \bibinfo {pages} {346} (\bibinfo {year} {2021})}\BibitemShut {NoStop}%
\bibitem [{\citenamefont {Li}\ \emph {et~al.}(2021)\citenamefont {Li}, \citenamefont {Jin}, \citenamefont {Wu}, \citenamefont {Chang}, \citenamefont {Wang}, \citenamefont {Shen}, \citenamefont {Yuan}, \citenamefont {Feshali}, \citenamefont {Paniccia}, \citenamefont {Vahala} \emph {et~al.}}]{li2021reaching}%
  \BibitemOpen
  \bibfield  {author} {\bibinfo {author} {\bibfnamefont {B.}~\bibnamefont {Li}}, \bibinfo {author} {\bibfnamefont {W.}~\bibnamefont {Jin}}, \bibinfo {author} {\bibfnamefont {L.}~\bibnamefont {Wu}}, \bibinfo {author} {\bibfnamefont {L.}~\bibnamefont {Chang}}, \bibinfo {author} {\bibfnamefont {H.}~\bibnamefont {Wang}}, \bibinfo {author} {\bibfnamefont {B.}~\bibnamefont {Shen}}, \bibinfo {author} {\bibfnamefont {Z.}~\bibnamefont {Yuan}}, \bibinfo {author} {\bibfnamefont {A.}~\bibnamefont {Feshali}}, \bibinfo {author} {\bibfnamefont {M.}~\bibnamefont {Paniccia}}, \bibinfo {author} {\bibfnamefont {K.~J.}\ \bibnamefont {Vahala}}, \emph {et~al.},\ }\bibfield  {title} {\bibinfo {title} {Reaching fiber-laser coherence in integrated photonics},\ }\href@noop {} {\bibfield  {journal} {\bibinfo  {journal} {Optics Letters}\ }\textbf {\bibinfo {volume} {46}},\ \bibinfo {pages} {5201} (\bibinfo {year} {2021})}\BibitemShut {NoStop}%
\bibitem [{\citenamefont {Sun}\ \emph {et~al.}(2024)\citenamefont {Sun}, \citenamefont {Wang}, \citenamefont {Liu}, \citenamefont {Harrington}, \citenamefont {Tabatabaei}, \citenamefont {Liu}, \citenamefont {Wang}, \citenamefont {Hanifi}, \citenamefont {Morgan}, \citenamefont {Jahanbozorgi} \emph {et~al.}}]{sun2024integrated}%
  \BibitemOpen
  \bibfield  {author} {\bibinfo {author} {\bibfnamefont {S.}~\bibnamefont {Sun}}, \bibinfo {author} {\bibfnamefont {B.}~\bibnamefont {Wang}}, \bibinfo {author} {\bibfnamefont {K.}~\bibnamefont {Liu}}, \bibinfo {author} {\bibfnamefont {M.~W.}\ \bibnamefont {Harrington}}, \bibinfo {author} {\bibfnamefont {F.}~\bibnamefont {Tabatabaei}}, \bibinfo {author} {\bibfnamefont {R.}~\bibnamefont {Liu}}, \bibinfo {author} {\bibfnamefont {J.}~\bibnamefont {Wang}}, \bibinfo {author} {\bibfnamefont {S.}~\bibnamefont {Hanifi}}, \bibinfo {author} {\bibfnamefont {J.~S.}\ \bibnamefont {Morgan}}, \bibinfo {author} {\bibfnamefont {M.}~\bibnamefont {Jahanbozorgi}}, \emph {et~al.},\ }\bibfield  {title} {\bibinfo {title} {Integrated optical frequency division for microwave and mm{W}ave generation},\ }\href@noop {} {\bibfield  {journal} {\bibinfo  {journal} {Nature}\ ,\ \bibinfo {pages} {1}} (\bibinfo {year} {2024})}\BibitemShut {NoStop}%
\bibitem [{\citenamefont {Idjadi}\ \emph {et~al.}(2024)\citenamefont {Idjadi}, \citenamefont {Kim},\ and\ \citenamefont {Fontaine}}]{idjadi2024modulation}%
  \BibitemOpen
  \bibfield  {author} {\bibinfo {author} {\bibfnamefont {M.~H.}\ \bibnamefont {Idjadi}}, \bibinfo {author} {\bibfnamefont {K.}~\bibnamefont {Kim}},\ and\ \bibinfo {author} {\bibfnamefont {N.~K.}\ \bibnamefont {Fontaine}},\ }\bibfield  {title} {\bibinfo {title} {Modulation-free laser stabilization technique using integrated cavity-coupled {M}ach-{Z}ehnder interferometer},\ }\href@noop {} {\bibfield  {journal} {\bibinfo  {journal} {Nature Communications}\ }\textbf {\bibinfo {volume} {15}},\ \bibinfo {pages} {1922} (\bibinfo {year} {2024})}\BibitemShut {NoStop}%
\bibitem [{\citenamefont {He}\ \emph {et~al.}(2024)\citenamefont {He}, \citenamefont {Cheng}, \citenamefont {Wang}, \citenamefont {Zhang}, \citenamefont {Meade}, \citenamefont {Vahala}, \citenamefont {Zhang},\ and\ \citenamefont {Li}}]{he2024chip}%
  \BibitemOpen
  \bibfield  {author} {\bibinfo {author} {\bibfnamefont {Y.}~\bibnamefont {He}}, \bibinfo {author} {\bibfnamefont {L.}~\bibnamefont {Cheng}}, \bibinfo {author} {\bibfnamefont {H.}~\bibnamefont {Wang}}, \bibinfo {author} {\bibfnamefont {Y.}~\bibnamefont {Zhang}}, \bibinfo {author} {\bibfnamefont {R.}~\bibnamefont {Meade}}, \bibinfo {author} {\bibfnamefont {K.}~\bibnamefont {Vahala}}, \bibinfo {author} {\bibfnamefont {M.}~\bibnamefont {Zhang}},\ and\ \bibinfo {author} {\bibfnamefont {J.}~\bibnamefont {Li}},\ }\bibfield  {title} {\bibinfo {title} {Chip-scale high-performance photonic microwave oscillator},\ }\href@noop {} {\bibfield  {journal} {\bibinfo  {journal} {arXiv preprint arXiv:2402.16229}\ } (\bibinfo {year} {2024})}\BibitemShut {NoStop}%
\bibitem [{\citenamefont {Wu}\ \emph {et~al.}(2020)\citenamefont {Wu}, \citenamefont {Wang}, \citenamefont {Yang}, \citenamefont {Ji}, \citenamefont {Shen}, \citenamefont {Bao}, \citenamefont {Gao},\ and\ \citenamefont {Vahala}}]{wu2020greater}%
  \BibitemOpen
  \bibfield  {author} {\bibinfo {author} {\bibfnamefont {L.}~\bibnamefont {Wu}}, \bibinfo {author} {\bibfnamefont {H.}~\bibnamefont {Wang}}, \bibinfo {author} {\bibfnamefont {Q.}~\bibnamefont {Yang}}, \bibinfo {author} {\bibfnamefont {Q.-x.}\ \bibnamefont {Ji}}, \bibinfo {author} {\bibfnamefont {B.}~\bibnamefont {Shen}}, \bibinfo {author} {\bibfnamefont {C.}~\bibnamefont {Bao}}, \bibinfo {author} {\bibfnamefont {M.}~\bibnamefont {Gao}},\ and\ \bibinfo {author} {\bibfnamefont {K.}~\bibnamefont {Vahala}},\ }\bibfield  {title} {\bibinfo {title} {Greater than one billion {Q} factor for on-chip microresonators},\ }\href@noop {} {\bibfield  {journal} {\bibinfo  {journal} {Optics Letters}\ }\textbf {\bibinfo {volume} {45}},\ \bibinfo {pages} {5129} (\bibinfo {year} {2020})}\BibitemShut {NoStop}%
\bibitem [{\citenamefont {Jin}\ \emph {et~al.}(2024)\citenamefont {Jin}, \citenamefont {Xie}, \citenamefont {Zhang}, \citenamefont {Hou}, \citenamefont {Zhang}, \citenamefont {Zhang}, \citenamefont {Chang}, \citenamefont {Gong},\ and\ \citenamefont {Yang}}]{jin2024microresonator}%
  \BibitemOpen
  \bibfield  {author} {\bibinfo {author} {\bibfnamefont {X.}~\bibnamefont {Jin}}, \bibinfo {author} {\bibfnamefont {Z.}~\bibnamefont {Xie}}, \bibinfo {author} {\bibfnamefont {X.}~\bibnamefont {Zhang}}, \bibinfo {author} {\bibfnamefont {H.}~\bibnamefont {Hou}}, \bibinfo {author} {\bibfnamefont {F.}~\bibnamefont {Zhang}}, \bibinfo {author} {\bibfnamefont {X.}~\bibnamefont {Zhang}}, \bibinfo {author} {\bibfnamefont {L.}~\bibnamefont {Chang}}, \bibinfo {author} {\bibfnamefont {Q.}~\bibnamefont {Gong}},\ and\ \bibinfo {author} {\bibfnamefont {Q.-F.}\ \bibnamefont {Yang}},\ }\bibfield  {title} {\bibinfo {title} {Microresonator-referenced soliton microcombs with zeptosecond-level timing noise},\ }\href@noop {} {\bibfield  {journal} {\bibinfo  {journal} {arXiv preprint arXiv:2401.12760}\ } (\bibinfo {year} {2024})}\BibitemShut {NoStop}%
\bibitem [{\citenamefont {Lihachev}\ \emph {et~al.}(2022)\citenamefont {Lihachev}, \citenamefont {Weng}, \citenamefont {Liu}, \citenamefont {Chang}, \citenamefont {Guo}, \citenamefont {He}, \citenamefont {Wang}, \citenamefont {Anderson}, \citenamefont {Liu}, \citenamefont {Bowers} \emph {et~al.}}]{lihachev2022platicon}%
  \BibitemOpen
  \bibfield  {author} {\bibinfo {author} {\bibfnamefont {G.}~\bibnamefont {Lihachev}}, \bibinfo {author} {\bibfnamefont {W.}~\bibnamefont {Weng}}, \bibinfo {author} {\bibfnamefont {J.}~\bibnamefont {Liu}}, \bibinfo {author} {\bibfnamefont {L.}~\bibnamefont {Chang}}, \bibinfo {author} {\bibfnamefont {J.}~\bibnamefont {Guo}}, \bibinfo {author} {\bibfnamefont {J.}~\bibnamefont {He}}, \bibinfo {author} {\bibfnamefont {R.~N.}\ \bibnamefont {Wang}}, \bibinfo {author} {\bibfnamefont {M.~H.}\ \bibnamefont {Anderson}}, \bibinfo {author} {\bibfnamefont {Y.}~\bibnamefont {Liu}}, \bibinfo {author} {\bibfnamefont {J.~E.}\ \bibnamefont {Bowers}}, \emph {et~al.},\ }\bibfield  {title} {\bibinfo {title} {Platicon microcomb generation using laser self-injection locking},\ }\href@noop {} {\bibfield  {journal} {\bibinfo  {journal} {Nature communications}\ }\textbf {\bibinfo {volume} {13}},\ \bibinfo {pages} {1771} (\bibinfo {year} {2022})}\BibitemShut {NoStop}%
\bibitem [{\citenamefont {Xiang}\ \emph {et~al.}(2023)\citenamefont {Xiang}, \citenamefont {Jin}, \citenamefont {Terra}, \citenamefont {Dong}, \citenamefont {Wang}, \citenamefont {Wu}, \citenamefont {Guo}, \citenamefont {Morin}, \citenamefont {Hughes}, \citenamefont {Peters} \emph {et~al.}}]{xiang20233d}%
  \BibitemOpen
  \bibfield  {author} {\bibinfo {author} {\bibfnamefont {C.}~\bibnamefont {Xiang}}, \bibinfo {author} {\bibfnamefont {W.}~\bibnamefont {Jin}}, \bibinfo {author} {\bibfnamefont {O.}~\bibnamefont {Terra}}, \bibinfo {author} {\bibfnamefont {B.}~\bibnamefont {Dong}}, \bibinfo {author} {\bibfnamefont {H.}~\bibnamefont {Wang}}, \bibinfo {author} {\bibfnamefont {L.}~\bibnamefont {Wu}}, \bibinfo {author} {\bibfnamefont {J.}~\bibnamefont {Guo}}, \bibinfo {author} {\bibfnamefont {T.~J.}\ \bibnamefont {Morin}}, \bibinfo {author} {\bibfnamefont {E.}~\bibnamefont {Hughes}}, \bibinfo {author} {\bibfnamefont {J.}~\bibnamefont {Peters}}, \emph {et~al.},\ }\bibfield  {title} {\bibinfo {title} {3{D} integration enables ultralow-noise isolator-free lasers in silicon photonics},\ }\href@noop {} {\bibfield  {journal} {\bibinfo  {journal} {Nature}\ }\textbf {\bibinfo {volume} {620}},\ \bibinfo {pages} {78} (\bibinfo {year} {2023})}\BibitemShut {NoStop}%
\bibitem [{\citenamefont {Liu}\ \emph {et~al.}(2022)\citenamefont {Liu}, \citenamefont {Jin}, \citenamefont {Cheng}, \citenamefont {Chauhan}, \citenamefont {Puckett}, \citenamefont {Nelson}, \citenamefont {Behunin}, \citenamefont {Rakich},\ and\ \citenamefont {Blumenthal}}]{liu2022ultralow}%
  \BibitemOpen
  \bibfield  {author} {\bibinfo {author} {\bibfnamefont {K.}~\bibnamefont {Liu}}, \bibinfo {author} {\bibfnamefont {N.}~\bibnamefont {Jin}}, \bibinfo {author} {\bibfnamefont {H.}~\bibnamefont {Cheng}}, \bibinfo {author} {\bibfnamefont {N.}~\bibnamefont {Chauhan}}, \bibinfo {author} {\bibfnamefont {M.~W.}\ \bibnamefont {Puckett}}, \bibinfo {author} {\bibfnamefont {K.~D.}\ \bibnamefont {Nelson}}, \bibinfo {author} {\bibfnamefont {R.~O.}\ \bibnamefont {Behunin}}, \bibinfo {author} {\bibfnamefont {P.~T.}\ \bibnamefont {Rakich}},\ and\ \bibinfo {author} {\bibfnamefont {D.~J.}\ \bibnamefont {Blumenthal}},\ }\bibfield  {title} {\bibinfo {title} {Ultralow 0.034 d{B}/m loss wafer-scale integrated photonics realizing 720 million {Q} and 380 $\mu$w threshold brillouin lasing},\ }\href@noop {} {\bibfield  {journal} {\bibinfo  {journal} {Optics letters}\ }\textbf {\bibinfo {volume} {47}},\ \bibinfo {pages} {1855} (\bibinfo {year} {2022})}\BibitemShut {NoStop}%
\bibitem [{\citenamefont {Liang}\ and\ \citenamefont {Liu}(2023)}]{liang2023compact}%
  \BibitemOpen
  \bibfield  {author} {\bibinfo {author} {\bibfnamefont {W.}~\bibnamefont {Liang}}\ and\ \bibinfo {author} {\bibfnamefont {Y.}~\bibnamefont {Liu}},\ }\bibfield  {title} {\bibinfo {title} {Compact sub-hertz linewidth laser enabled by self-injection lock to a sub-milliliter {FP} cavity},\ }\href@noop {} {\bibfield  {journal} {\bibinfo  {journal} {Optics Letters}\ }\textbf {\bibinfo {volume} {48}},\ \bibinfo {pages} {1323} (\bibinfo {year} {2023})}\BibitemShut {NoStop}%
\bibitem [{\citenamefont {Dahmani}\ \emph {et~al.}(1987)\citenamefont {Dahmani}, \citenamefont {Hollberg},\ and\ \citenamefont {Drullinger}}]{dahmani1987frequency}%
  \BibitemOpen
  \bibfield  {author} {\bibinfo {author} {\bibfnamefont {B.}~\bibnamefont {Dahmani}}, \bibinfo {author} {\bibfnamefont {L.}~\bibnamefont {Hollberg}},\ and\ \bibinfo {author} {\bibfnamefont {R.}~\bibnamefont {Drullinger}},\ }\bibfield  {title} {\bibinfo {title} {Frequency stabilization of semiconductor lasers by resonant optical feedback},\ }\href@noop {} {\bibfield  {journal} {\bibinfo  {journal} {Optics letters}\ }\textbf {\bibinfo {volume} {12}},\ \bibinfo {pages} {876} (\bibinfo {year} {1987})}\BibitemShut {NoStop}%
\bibitem [{\citenamefont {Laurent}\ \emph {et~al.}(1989)\citenamefont {Laurent}, \citenamefont {Clairon},\ and\ \citenamefont {Breant}}]{laurent1989frequency}%
  \BibitemOpen
  \bibfield  {author} {\bibinfo {author} {\bibfnamefont {P.}~\bibnamefont {Laurent}}, \bibinfo {author} {\bibfnamefont {A.}~\bibnamefont {Clairon}},\ and\ \bibinfo {author} {\bibfnamefont {C.}~\bibnamefont {Breant}},\ }\bibfield  {title} {\bibinfo {title} {Frequency noise analysis of optically self-locked diode lasers},\ }\href@noop {} {\bibfield  {journal} {\bibinfo  {journal} {IEEE Journal of Quantum Electronics}\ }\textbf {\bibinfo {volume} {25}},\ \bibinfo {pages} {1131} (\bibinfo {year} {1989})}\BibitemShut {NoStop}%
\bibitem [{\citenamefont {Hjelme}\ \emph {et~al.}(1991)\citenamefont {Hjelme}, \citenamefont {Mickelson},\ and\ \citenamefont {Beausoleil}}]{hjelme1991semiconductor}%
  \BibitemOpen
  \bibfield  {author} {\bibinfo {author} {\bibfnamefont {D.~R.}\ \bibnamefont {Hjelme}}, \bibinfo {author} {\bibfnamefont {A.~R.}\ \bibnamefont {Mickelson}},\ and\ \bibinfo {author} {\bibfnamefont {R.~G.}\ \bibnamefont {Beausoleil}},\ }\bibfield  {title} {\bibinfo {title} {Semiconductor laser stabilization by external optical feedback},\ }\href@noop {} {\bibfield  {journal} {\bibinfo  {journal} {IEEE journal of quantum electronics}\ }\textbf {\bibinfo {volume} {27}},\ \bibinfo {pages} {352} (\bibinfo {year} {1991})}\BibitemShut {NoStop}%
\bibitem [{\citenamefont {Savchenkov}\ \emph {et~al.}(2024)\citenamefont {Savchenkov}, \citenamefont {Zhang}, \citenamefont {Iltchenko},\ and\ \citenamefont {Matsko}}]{savchenkov2024robust}%
  \BibitemOpen
  \bibfield  {author} {\bibinfo {author} {\bibfnamefont {A.}~\bibnamefont {Savchenkov}}, \bibinfo {author} {\bibfnamefont {W.}~\bibnamefont {Zhang}}, \bibinfo {author} {\bibfnamefont {V.}~\bibnamefont {Iltchenko}},\ and\ \bibinfo {author} {\bibfnamefont {A.}~\bibnamefont {Matsko}},\ }\bibfield  {title} {\bibinfo {title} {Robust self-injection locking to a non-confocal monolithic {F}abry--{P}erot cavity},\ }\href@noop {} {\bibfield  {journal} {\bibinfo  {journal} {Optics Letters}\ }\textbf {\bibinfo {volume} {49}},\ \bibinfo {pages} {1520} (\bibinfo {year} {2024})}\BibitemShut {NoStop}%
\bibitem [{\citenamefont {Marpaung}\ \emph {et~al.}(2013)\citenamefont {Marpaung}, \citenamefont {Morrison}, \citenamefont {Pant}, \citenamefont {Roeloffzen}, \citenamefont {Leinse}, \citenamefont {Hoekman}, \citenamefont {Heideman},\ and\ \citenamefont {Eggleton}}]{marpaung2013si}%
  \BibitemOpen
  \bibfield  {author} {\bibinfo {author} {\bibfnamefont {D.}~\bibnamefont {Marpaung}}, \bibinfo {author} {\bibfnamefont {B.}~\bibnamefont {Morrison}}, \bibinfo {author} {\bibfnamefont {R.}~\bibnamefont {Pant}}, \bibinfo {author} {\bibfnamefont {C.}~\bibnamefont {Roeloffzen}}, \bibinfo {author} {\bibfnamefont {A.}~\bibnamefont {Leinse}}, \bibinfo {author} {\bibfnamefont {M.}~\bibnamefont {Hoekman}}, \bibinfo {author} {\bibfnamefont {R.}~\bibnamefont {Heideman}},\ and\ \bibinfo {author} {\bibfnamefont {B.~J.}\ \bibnamefont {Eggleton}},\ }\bibfield  {title} {\bibinfo {title} {Si3{N}4 ring resonator-based microwave photonic notch filter with an ultrahigh peak rejection},\ }\href@noop {} {\bibfield  {journal} {\bibinfo  {journal} {Optics express}\ }\textbf {\bibinfo {volume} {21}},\ \bibinfo {pages} {23286} (\bibinfo {year} {2013})}\BibitemShut {NoStop}%
\bibitem [{\citenamefont {Herrmann}\ \emph {et~al.}(2022)\citenamefont {Herrmann}, \citenamefont {Ansari}, \citenamefont {Wang}, \citenamefont {Witmer}, \citenamefont {Fan},\ and\ \citenamefont {Safavi-Naeini}}]{herrmann2022mirror}%
  \BibitemOpen
  \bibfield  {author} {\bibinfo {author} {\bibfnamefont {J.~F.}\ \bibnamefont {Herrmann}}, \bibinfo {author} {\bibfnamefont {V.}~\bibnamefont {Ansari}}, \bibinfo {author} {\bibfnamefont {J.}~\bibnamefont {Wang}}, \bibinfo {author} {\bibfnamefont {J.~D.}\ \bibnamefont {Witmer}}, \bibinfo {author} {\bibfnamefont {S.}~\bibnamefont {Fan}},\ and\ \bibinfo {author} {\bibfnamefont {A.~H.}\ \bibnamefont {Safavi-Naeini}},\ }\bibfield  {title} {\bibinfo {title} {Mirror symmetric on-chip frequency circulation of light},\ }\href@noop {} {\bibfield  {journal} {\bibinfo  {journal} {Nature Photonics}\ }\textbf {\bibinfo {volume} {16}},\ \bibinfo {pages} {603} (\bibinfo {year} {2022})}\BibitemShut {NoStop}%
\bibitem [{\citenamefont {Cheng}\ \emph {et~al.}(2024)\citenamefont {Cheng}, \citenamefont {Zhou}, \citenamefont {Ruesink}, \citenamefont {Pavlovich}, \citenamefont {Gertler}, \citenamefont {Starbuck}, \citenamefont {Leenheer}, \citenamefont {Pomerene}, \citenamefont {Trotter}, \citenamefont {Dallo} \emph {et~al.}}]{cheng2024terahertz}%
  \BibitemOpen
  \bibfield  {author} {\bibinfo {author} {\bibfnamefont {H.}~\bibnamefont {Cheng}}, \bibinfo {author} {\bibfnamefont {Y.}~\bibnamefont {Zhou}}, \bibinfo {author} {\bibfnamefont {F.}~\bibnamefont {Ruesink}}, \bibinfo {author} {\bibfnamefont {M.}~\bibnamefont {Pavlovich}}, \bibinfo {author} {\bibfnamefont {S.}~\bibnamefont {Gertler}}, \bibinfo {author} {\bibfnamefont {A.~L.}\ \bibnamefont {Starbuck}}, \bibinfo {author} {\bibfnamefont {A.~J.}\ \bibnamefont {Leenheer}}, \bibinfo {author} {\bibfnamefont {A.~T.}\ \bibnamefont {Pomerene}}, \bibinfo {author} {\bibfnamefont {D.~C.}\ \bibnamefont {Trotter}}, \bibinfo {author} {\bibfnamefont {C.}~\bibnamefont {Dallo}}, \emph {et~al.},\ }\bibfield  {title} {\bibinfo {title} {A terahertz bandwidth nonmagnetic isolator},\ }\href@noop {} {\bibfield  {journal} {\bibinfo  {journal} {arXiv preprint arXiv:2403.10628}\ } (\bibinfo {year} {2024})}\BibitemShut {NoStop}%
\end{thebibliography}%

\noindent \textbf{Corresponding authors}: \href{mailto:peter.rakich@yale.edu}{peter.rakich@yale.edu} and \href{mailto:haotian.cheng@yale.edu}{haotian.cheng@yale.edu}\vspace{6pt}

\noindent \textbf{Acknowledgment}: 
We thank Charles McLemore, Takuma Nakamura and William Groman for useful discussion and feedback.

This work was supported by Defense Advanced Research Projects Agency (DARPA) under award number HR0011-22-2-0009 as well as U.S. Department of Energy (DoE) under award number DE-SC0019406 and the National Science Foundation (NSF) under award number 2137740. Any opinions, findings, and conclusions or recommendations expressed in this publication are those of the authors and do not necessarily reflect the views of DARPA, DoE, and NSF.

\noindent \textbf{Author Contributions}:

H.C., C.X., K.J.V., F.Q., S.A.D., J. B. and P.T.R. led the project and conceived the physics and experiment.

H.C., C.X. designed SIL/PDH interface circuits.

N.J., and H.C. fabricated and packaged \textmu FP cavity.

H.C., I.K. ,N.J. and M.H. measured SIL laser.

J.G., and J.P. fabricated the SIL interface circuit.

H.C. fabricated and measured PDH interface circuits.

H.C., C.X., N.J. and P.T.R. wrote the paper with input from all authors.

All authors contributed to the design and discussion of the results.

\noindent \textbf{Data Availability}
Data sets generated during the current study are available from the corresponding author on reasonable request.
\end{document}